\documentclass[prl,letterpaper,twocolumn,final,superscriptaddress,floatfix]{revtex4-1}

\usepackage{graphicx,psfrag,amsmath,amssymb,amsfonts,bbm,latexsym,color,dcolumn,bm,mathbbol}
\usepackage[framemethod=tikz]{mdframed}
\usepackage{hyperref}
\usepackage{xfrac} 
\usepackage{ifthen}

\usepackage{amsmath,amssymb}             
\usepackage[french, english]{babel}

\newcommand{\modif}[1]{ {{#1}} }  

\usepackage{enumerate}
\usepackage{color}
\begin{document}

\title{Unsual ionic transport and fluctuation in indivual carbon nanotube : the role of surface conduction}
\title{Scaling behavior of surface conduction and fluctuations of ionic transport in individual carbon nanotube}
\title{Scaling behavior for ionic transport and current fluctuations in individual carbon nanotube}
\title{Scaling behavior for ionic transport and its fluctuations in individual carbon nanotubes}

\author{Eleonora Secchi}
\affiliation{Laboratoire de Physique Statistique de l'Ecole Normale Sup\'erieure, UMR 8550, 24 Rue Lhomond 75005 Paris, France}
\author{Antoine Nigu\`es}
\affiliation{Laboratoire de Physique Statistique de l'Ecole Normale Sup\'erieure, UMR 8550, 24 Rue Lhomond 75005 Paris, France}
\author{Laetitia Jubin}
\affiliation{Laboratoire de Physique Statistique de l'Ecole Normale Sup\'erieure, UMR 8550, 24 Rue Lhomond 75005 Paris, France}
\author{Alessandro Siria}
\affiliation{Laboratoire de Physique Statistique de l'Ecole Normale Sup\'erieure, UMR 8550, 24 Rue Lhomond 75005 Paris, France}
\author{Lyd\'eric Bocquet}
\affiliation{Laboratoire de Physique Statistique de l'Ecole Normale Sup\'erieure, UMR 8550, 24 Rue Lhomond 75005 Paris, France}


\begin{abstract}
In this letter we perform an experimental study of ionic transport and current fluctuations inside individual Carbon Nanotubes (CNT).
The conductance exhibits a power law behavior at low salinity, with an exponent close to 1/3 versus the salt concentration in this regime. This  behavior is rationalized in terms of a salinity dependent surface charge, which is accounted for on the basis of a model for hydroxide adsorption at the (hydrophobic) carbon surface. This is in contrast to boron nitride nanotubes which exhibit a constant surface conductance. Further we measure the low frequency noise of the ionic current in CNT and show that the amplitude of the noise scales with the surface charge, with data collapsing on a master curve for the various studied CNT at a given pH.
\end{abstract}

\maketitle


Transport of fluids at nanoscale remains to a large extent a virgin territory. Over the recent years, new phenomena have been
unveiled such as fast flows  \cite{Bakajin, Hinds, Whitby,Lindsay} or peculiar ion transport  in carbon nanotubes \cite{Strano}, large osmotic power in boron nitride nanotubes \cite{Siria}, or high permeation across nanoporous graphene and graphene oxides \cite{Karnik2013,Geim2014,Park2014}. Many of these phenomena remain to be rationalized \cite{Bocquet2010,Bocquet2014}. 
While the field has been explored exhaustively on the theoretical and numerical side, there is 
still a lack of experimental output, as studies in this domain are very challenging. However a systematic understanding of fluidic transport within nanochannels, and in particular the somewhat mysterious carbon materials, is a prerequisite to gain fundamental insights into the mechanisms at play at the nanoscales.  A lot of hope has indeed been raised by the fluidic properties of these materials with impact on societal questions like desalination and energy harvesting, and it is accordingly crucial to pinpoint the physical origin of their specific behavior.


In this paper,  we explore ionic transport inside individual carbon nanotubes (CNT) of various sizes, typically in the ten of nanometers range. We focus in particular on the ionic conductance and its dependence on salt concentration, as well as on the fluctuations of the ionic current. We report an 'unusual' scaling behavior of the conductance at low salt concentration, which can be interpreted in terms of hydroxide adsorption on the carbon surface. Further the measurements of the current noise highlight an intimate dependence of the noise amplitude on the surface charge, suggesting that surface adsorption plays a key role in the low frequency behavior of ionic transport.
Results are shown to be strongly different to the response of boron nitride nanotubes (BNNT), which exhibit the same crystallography but radically different electronic properties. 


\begin{figure}[h]
\includegraphics[width=\columnwidth]{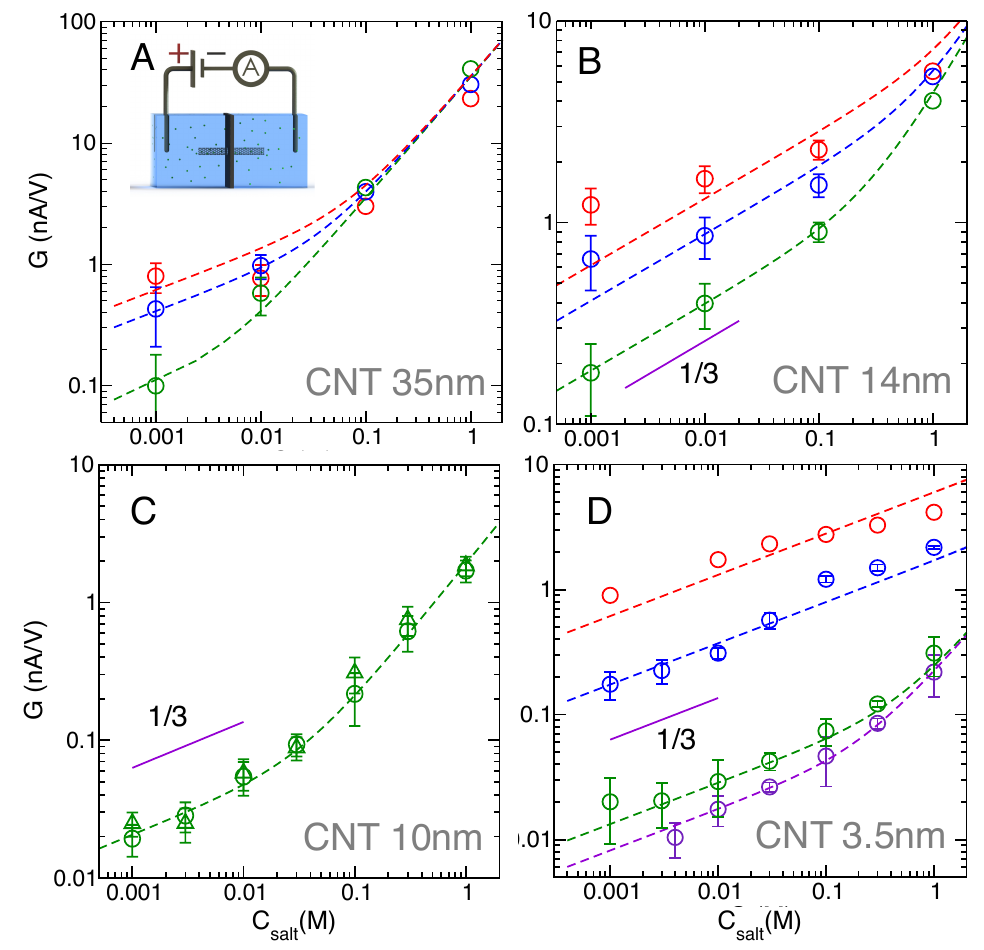}
\caption{Conductance $G=I/\Delta V$ measured inside single carbon nanotubes with various geometrical characteristics and pH.  
(A): $\{R,L\}=\{35$ nm$, 1500$ nm$\}$, pH=6, 9, 10 from bottom to top; (B): $\{R,L\}=\{14$ nm$, 2000$ nm$\}$, pH=6, 9, 10 from bottom to top; (C): $\{R,L\}=\{10$ nm$, 2500$ nm$\}$, pH=4; (D): $\{R,L\}=\{3.5$ nm$, 3000$ nm$\}$, pH=4, 6, 8, 10 from bottom to top. 
The dashed lines are predictions according to the model in the text, $G= 2 e^2\,\mu\, {\pi R^2\over L}  \sqrt{\rho_s^2  +  {\Sigma^2/ R^2}} $, with a surface charge scaling with salt concentration $C_{\rm salt}$ as $\Sigma \propto C_{\rm salt}^{1/3}$ ($C_{\rm salt}$ in Mol.L$^{-1}$ and $\rho_s=6.02\, 10^{26} \times C_s$ the numerical density in m$^{-3}$). See text and Supplemental Material for details. 
}
\label{fig:Figure1}
\end{figure}

{\it Individual nanotube and Experimental setup --}
The individual trans-membrane nanotube device consists of an individual nanotube inserted in the hole of a pierced Silicon Nitride membrane. The methodology to fabricate these nanosystems is similar to that reported in Ref. \cite{Siria}, here extended to carbon nanotubes. Briefly, a single multiwall nanotube is inserted inside a $\sim$ 200 nm hole, previously pierced using Focus Ion Beam (FIB). The insertion is performed using a home-made nanomanipulator, which consists of  piezo-inertial step motors, inside a Scaning Electronic Microscope (SEM). Sealing is obtained by cracking naphtalene locally under the electronic beam. All details of this procedure, as well as the benchmarking steps and cross-checks, follow those described in detail in the Supplemental Material of \cite{Siria}. 
{Carbon nanotubes used in this study are commercially available from Sigma Aldrich (ref 659258)}.
The transmembrane nanotube is then squeezed between two macroscopic fluid reservoirs containing potassium chloride (KCl) solutions of various concentrations; Ag/AgCl electrodes inside the reservoirs were connected to an external patch-clamp amplifier for the electrical measurements with a resolution in the tens of picoampere for a 5 kHz sampling rate.

Here we performed experiments with five carbon nanotubes - radius $R\simeq 35$ nm and length $L\simeq 1500$ nm, radius $R\simeq 14$ nm and length $L\simeq 2000$ nm, radius $R\simeq 10$ nm and length $L\simeq 2500$ nm,  radius $R\simeq 7$ nm and length $L\simeq 1000$ nm, radius $R\simeq 3.5$ nm and length $L\simeq 3000$ nm -. Two boron nitride nanotubes, with  $R\simeq 15$ nm and  $L=800$nm, and  $R\simeq 25$ nm  length $L\simeq 500$ nm were also considered for comparison.

{\it Ionic transport and scaling behavior of the conductance --}
We first measure the ionic current induced by a voltage drop.
 As for the BNNTs in \cite{Siria}, the current varies linearly with the voltage in all conditions (not shown).  Fig.~\ref{fig:Figure1} then reports the values for the conductance  $G=I/\Delta V$ versus the salt concentration for the various CNTs explored. A striking observation on Fig.~\ref{fig:Figure1} is that the conductance in CNTs does not exhibit a saturation at low salt concentration, { as expected for a nanochannel with a constant surface charge \cite{Bocquet2010} and exhibited by BNNTs, see Ref. \cite{Siria}. Rather a scaling behavior of the conductance with the salt concentration is exhibited at low salt concentration, in the form
\begin{equation}
G \sim C_{\rm salt}^{\alpha}
\label{scale}
\end{equation}
with an exponent  close to $\alpha =1/3$. This result echoes similar observations by Lindsay {\it et al. }\cite{Lindsay,Lindsay2011} and Forniasero {\it et al.} \cite{Forniasero} in CNTs. 
There is up to now no explanation for the origin of this behavior. 
\modif{Descriptions in terms of fixed surface charge or fixed potential at the walls are not able to account for this behavior. Also charged carboxylic groups localized at the CNT mouth, as put forward in \cite{Bakajin2008} to explain charge exclusion, are insufficient to explain a finite surface conduction along the CNT.}

Here we propose a possible interpretation for this behavior in terms of hydroxide adsorption at the surface of the (rather hydrophobic) carbon nanotubes, see Supplemental Information. Adsorption of ions at hydrophobic surface has been long suggested and many measurements suggest a preferential adsorption of hydroxide ions at surfaces at neutral and alkaline pH \cite{Zimmermann}, even though the origin is still debated. 
We accordingly assume that $OH^-$ ions have a preferred energy of adsorption, say ${\cal U}_{ads}$, at the carbon surface. In the context of   a standard charge-regulation model \cite{Israel}, the chemical potential of the $OH^-$ ions at the surface writes 
$\mu_i= k_BT \log( [OH^-]_s \lambda^2)- e V_s + {\cal U}_{ads}$ where the index $s$ denotes the value at the CNT surface and $V_s$ is the electrostatic potential ($\lambda$ is a microscopic normalizing length). We neglect here exclusion effects between hydroxide ions: this will lead to a saturation of the hydroxide concentration at the surface to a maximum limiting value. 
The surface charge $\vert\Sigma\vert \equiv [OH^-]_s$ is fixed by the equilibrium condition, $\mu_i = \mu_i[{\rm  bulk}]=k_BT \log ([OH^-]_{\rm bulk}\lambda^3)$. This yields $\vert\Sigma\vert =k_A \times \exp[\phi_s]$, with $\phi_s= e V_s/k_BT$, $k_A={K_A \,\lambda\over K_e} \, 10^{pH}$, $K_e=10^{14}$ the dissociation constant of water and $K_A=\exp[-{\cal U}_{ads}/k_BT]$ a surface adsorption constant.

This equation is complemented by the non-linear Poisson-Boltzmann (PB) equation for the electrostatic potential inside the tube. A full analytical resolution is not possible and we reduce the discussion to its key ingredients. A first integration of the PB equation coupled to the surface boundary condition leads to a relationship between the surface potential $\phi_s$, center potential $\phi_0$ and the surface charge: $\cosh \phi_s-\cosh \phi_0= 2/(\kappa \ell_{GC})^2$, with $\ell_{GC}= (2\pi \ell_B \vert\Sigma\vert)^{-1}$ the Gouy-Chapmann length, $\ell_B=e^2/4\pi \epsilon k_BT$ the Bjerrum length and $\kappa^{-1}=1/\sqrt{8\pi \ell_B \rho_s}$ the Debye length; $\rho_s$ is the salt number density (in m$^{-3}$ \modif{-- while the equivalent salt concentration $C_{salt}$ is in mole per liter--}).
In the limit where the surface potential is large and dominates over the averaged (Donnan) potential in the tube, the PB equation leads to the simple relationship between the surface potential and surface charge: $\cosh\phi_s \approx 2 (\kappa \ell_{GC})^{-2}$. 
Combining with the previous charge regulation condition for the surface charge, $\exp[\phi_s]=\vert\Sigma\vert /k_A $ (with $\phi_s<0$ for a negative surface charge), we obtain accordingly the self consistent equation for $\Sigma$ as
${k_A\over 2 \vert\Sigma\vert} \approx {\pi\ell_B \Sigma^2 \over \rho_s}$. Altogether this predicts
\begin{equation}
\vert\Sigma\vert \approx \left({k_A\over 2\pi \ell_B} \rho_s\right)^{1/3}
\label{Sigma}
\end{equation}
Note that we cannot exclude {\it at this stage} a proper covalent bonding of $-OH$ groups at the carbon surface, as in the end the physical chemistry would lead  to a qualitatively similar dependence.

Since it is expected that at low salt concentration the conductance is proportional to the surface charge \cite{Bocquet2010}, this prediction therefore explains the experimental observations in Fig.~\ref{fig:Figure1} and the scaling in Eq.(\ref{scale}).
Going further, we use the following expression for the conductance, based on the Poisson-Nernst-Planck model for conduction, which is adapted for confined geometries in the context of a Donnan description \cite{Bocquet2010,Manghi2015}
\begin{equation}
G= 2 e^2\,\mu\, {\pi R^2\over L}  \sqrt{\rho_s^2  +  {\Sigma^2\over R^2}} 
\label{G}
\end{equation}
with $\rho_s$ the KCl concentration, $e$ the electronic charge, $\mu={1\over 2} (\mu_{K^+}+\mu_{CL^-})=4.8\cdot 10^{11}$ s$\cdot$kg$^{-1}$ the KCl mobility and $e\Sigma$ the surface charge density on the surface (in C/m$^2$). 
Note that in order to keep the description to its simplest form, we omit the electro-osmotic conductance (which scales like $\Sigma^2$ \cite{Bocquet2010} {and can be checked {\it a posteriori} to be negligeable in the present conditions).} We also do not include specific ion mobility effects at the surface \cite{Vino}.
We combine this expression with a density dependent charge $\Sigma(\rho_s)$, defined in line with Eq.(\ref{Sigma}) as $\vert\Sigma\vert \ell_B^2 = C_0 (\rho_s \ell_B^3)^{1/3}$, with $C_0$ a dimensionless, pH dependent prefactor.

As shown in Fig.~\ref{fig:Figure1}, this prediction is altogether in good agreement with the experimental data for the various tubes and pH investigated.
As shown in the Supplemental Material, the constant $C_0$ extracted from the fits is found to depend on pH as $C_0\sim 10^{\beta pH}$, with an exponent $\beta \approx 0.2-0.33 $ in fair agreement with the prediction $\vert \Sigma\vert \propto k_A^{1/3} \propto 10^{pH/3}$, see Eq.(\ref{Sigma}). Some slight dependence of $C_0$ on the tube diameter remains which we cannot explain at this stage. From the fit of $C_0$ versus pH, one may then get a {\it rough} estimate of the adsorption constant $K_A$, which gives an adsorption energy 
${\cal U}_{ads} \approx 0.6$ eV:  this value is substantially smaller than covalent adsorption energies \cite{Angelos} and  rather suggests  physisorption of hydroxide ions at the CNT surface. This is {\it a posteriori} consistent with our initial assumption for the reversible adsorption of hydroxide at the CNT surface.




\begin{figure}[h]
\includegraphics[width=\columnwidth]{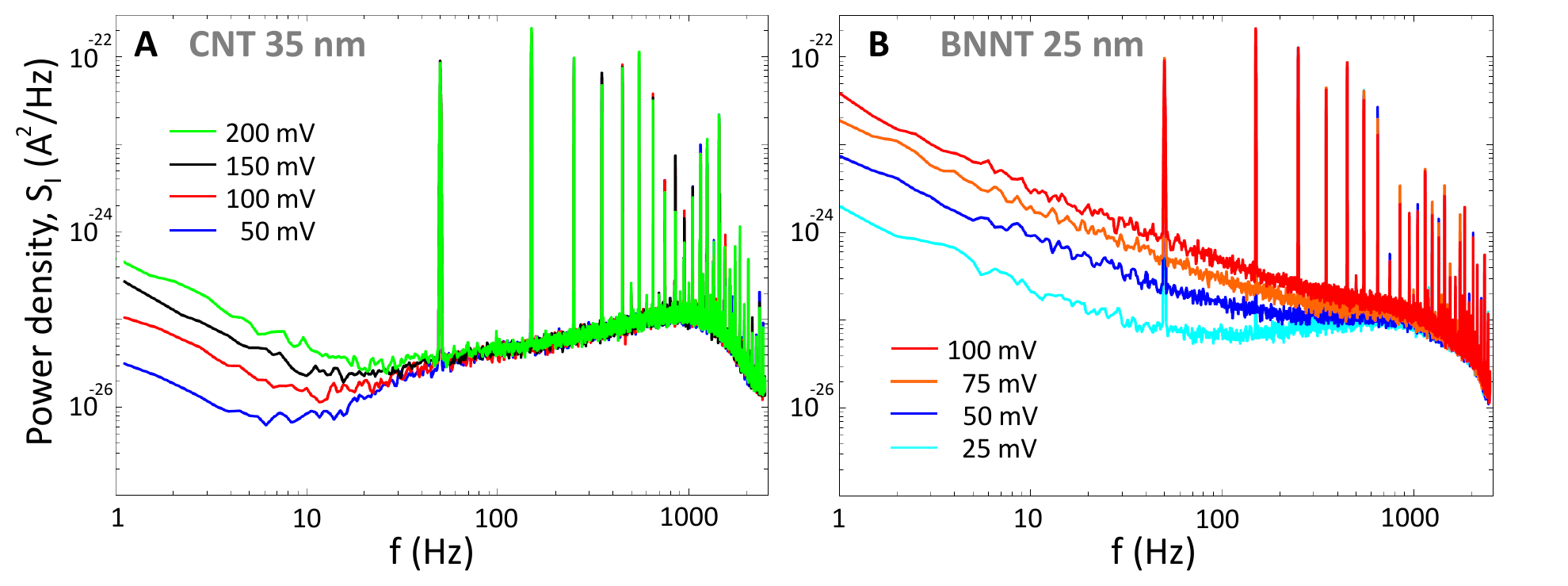}
\caption{Noise current power spectra for CNT with $\{R,L\}=\{35$ nm$, 1500$ nm$\}$ (Panel A) and BNNT with $\{R,L\}=\{25$ nm$, 500$ nm$\}$ (Panel B), measured at pH6 and at salt concentration equal to 1M, for different applied voltage. { Peaks in the noise at harmonics of 50 Hz are due to AC electric supply grid.}
}
\label{fig:Figure2}
\end{figure}

{\it Current fluctuations --}
The conductance measurements therefore show that the CNTs exhibit a quite unusual, but robust, surface behavior. In this context, we extend our investigations to the current noise. Specifically, we study
the fluctuations of the ionic current as a constant voltage drop is applied between the two sides of the tube. Experimentally the presence of a low frequency ($f\leq1$ kHz) pink noise in the ionic current has been repeatedly reported in solid state and biological nanopores \cite{Smeets2007,Siwy2009,Golovchenko,Tasserit2010,Benz1997}, but its origin has not found a satisfactory explanation up to now. Qualitatively such low frequency current noise is characterized by a power spectrum $S(f)\propto1/f^{\gamma}$ where $f$ is the frequency and with $\gamma\approx1$. A convenient way to describe such dependency is based on the long standing Hooge's empirical relation \cite{Hooge}:
\begin{equation}
S(f)=\frac{A_H}{f}=\frac{\alpha\, \bar I^2}{f}
\label{S}
\end{equation}
with $A_H$ the low frequency noise amplitude parameter and $\bar I$ the mean ionic current. According to Hooge, a (again empirical) observation is that  the noise amplitude is usually inversely dependent on the total number of charge carriers inside the channel $N_C$, $\alpha\propto1/N_C$, although there is no proper explanation for this behavior.

We have analyzed the fluctuations of the noise using {a Molecular Device Axopatch 200B. The ionic current is acquired with low pass filtering at a cut-off frequency of 5 kHz. The current is recorded and the power spectrum obtained with a home made LabView program}. In Fig.~\ref{fig:Figure2} we report the power spectra of the noise under various imposed voltage, for both a CNT and a BNNT. These measurements confirm the presence of a low frequency $1/f$ noise, whose amplitude $A_H$ increases with the imposed voltage.
We plot in Fig.~\ref{fig:Figure3}, the amplitude versus the mean current: as suggested by Eq.(\ref{S}), we find that the noise amplitude scales like the square of the mean (DC) current, $A_H\propto I^2$.

Varying the pH highlights a different behavior for the CNTs and the BNNT. As shown in Fig.~\ref{fig:Figure3}, the noise amplitude  $A_H$  for the CNT strongly varies with the salt concentration at low pH, while all results for $A_H$ follow the same functional dependence for the largest pH. The same behavior is observed for all CNTs with varying diameters, but with a noise amplitude which does depend on the CNT under scrutiny.
In contrast to CNTs, the noise amplitude for the BNNT remains insensitive to salt concentration whatever the pH. 


\begin{figure}[h]
\includegraphics[width=\columnwidth]{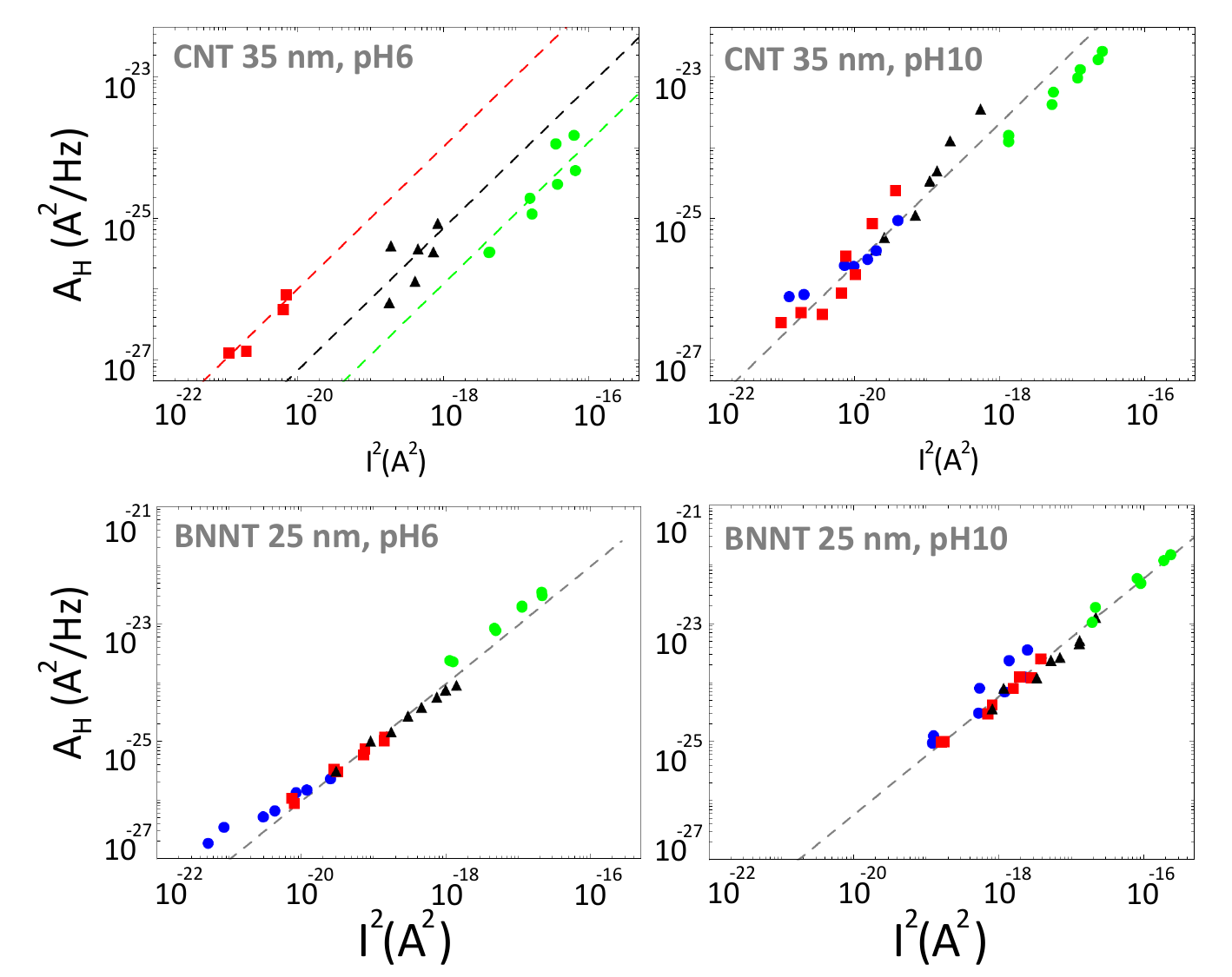}
\caption{Evolution of the parameter $A_H$ with the DC ionic current $I$ \modif{(log-log scales)}. The experiments have been performed on CNT with $\{R,L\}=\{35$ nm$, 1500$ nm$\}$ (Panel A and B) and BNNT with $\{R,L\}=\{25$ nm$, 500$ nm$\}$ (Panel C and D), for different values of pH (pH6, Panel A and C and pH10, Panel B and D; see Supplemental Material for the curves at pH9 ). The colors of the symbols correspond to various KCl concentration: $10^{-3}$ M (blue), $10^{-2}$ M (red), $10^{-1}$ M (black), $1$ M (green). \modif{The dashed lines have a slope 1 in log-log scales, highlighting the linear dependence of $A_H$ with $I^2$.}}
\label{fig:Figure3}
\end{figure}

While this behavior is somewhat complex, we show in Fig.~\ref{fig:Figure4} that  the data for the noise amplitude {rescaled by the { apparent} surface charge -- defined as above as $\Sigma_{app} \ell_B^2= C_0 (\rho_s \ell_B^3)^{1/3}$ with the parameter $C_0$ previously measured } -- do collapse on a single curve for all the investigated CNTs with various diameters, for pH 6 and 9. 
The fact that the noise should originate in process occurring at the surface was proposed for some time \cite{Golovchenko}. Here the  collapse highlighted experimentally for various systems strongly supports that the origin of the low frequency noise stems from process involving surface charge fluctuations.
\begin{figure}[h]
\includegraphics[width=\columnwidth]{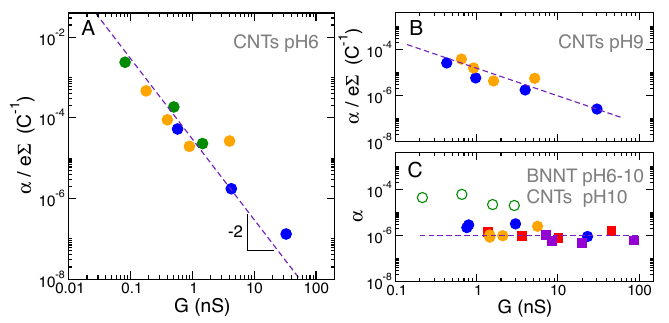}
\caption{Collapse of the noise amplitude rescaled by surface charge, $\alpha/e\Sigma $, as a function of the conductance $G$. The symbols are: CNT with $\{R,L\}=\{35$ nm$, 1500$ nm$\}$ (blue); CNT with $\{R,L\}=\{14$ nm$, 2000$ nm$\}$ (orange); CNT with $\{R,L\}=\{7$ nm$, 1000$ nm$\}$ (green). (A): results at pH6. The dashed line has a slope equal to -2,  $\alpha/\Sigma\propto 1/G^2$. (B): results at pH9. The dashed line has a slope equal to -1.2,  $\alpha/\Sigma\propto 1/G^{1.2}$. (C): $\alpha$ versus conductance for the three CNTs at pH10 and BNNT with $\{R,L\}=\{25$ nm$, 500$ nm$\}$ at pH6 (red) and pH10 (violet). }
\label{fig:Figure4}
\end{figure}

Simple arguments allow to get insights into the {\it amplitude} of {the low frequency} noise. First, one may assume that the low frequency noise fluctuations are linked to the conductance fluctuations, $\delta G$. Writing $\delta I= I-\bar I=
\delta G \times \Delta V$, with $\bar I=G \Delta V$ the mean current, then one gets for the noise spectrum $S(f)=\langle \delta I^2\rangle(f) =\langle \delta G^2\rangle(f) /   G^2\times  {\bar I}^2$. We rewrite formally this expression as 
 $S(f)= \langle \delta G^2\rangle /   G^2\times  {\bar I}^2\times {\cal F}(f)$, with ${\cal F}(f)$ a frequency dependent function which has the dimension of an {\it inverse frequency}. We will not discuss further the frequency dependence of ${\cal F}(f)$, which is found experimentally to scale like ${\cal F}(f)\sim 1/f$, { therefore following the longstanding Hooge's relationship}.  

We now rather focus on the noise amplitude.
Assuming that the noise takes its origin in the surface charge fluctuations, one may conjecture that
$\langle \delta G^2\rangle \propto e^2\langle \delta \Sigma^2\rangle$.
A crucial remark then is  that surface charge fluctuations are intimately related to the differential capacitance of the interface \cite{Limmer2013}, 
$e^2\langle \delta \Sigma^2\rangle = k_BT C_{\rm diff}$, 
with $C_{\rm diff}$ the differential capacitance defined as $C_{\rm diff}=  {\partial e\Sigma \over \partial V_s}$, with $V_s$ the surface 
electrostatic potential.
Within the approximated PB framework described above, the charge $e\Sigma$ is related to the potential via $\vert\Sigma\vert\approx {\kappa \over 4\pi \ell_B} \exp[-\phi_s/2]$, so that $C_{\rm diff}\simeq e^2 \vert \Sigma\vert/2k_BT$ in the present regime.
Gathering results,  the noise amplitude $A_H$ is predicted to scale as
\begin{equation}
A_H=\alpha\, \bar I^2 \propto {e\vert \Sigma\vert \over G^2} \times \bar I^2 
\end{equation}
Beyond the scaling in $\bar I^2$, this prediction therefore suggests that the noise amplitude $\alpha$ is directly proportional to the average surface charge: this is indeed in full agreement with the collapse of our experimental data in Fig.~\ref{fig:Figure4} for the various CNTs.
Even the scaling like $G^{-2}$ is recovered for pH6. For pH9, we still find a rescaling with the surface charge $e\Sigma$ but measure a different exponent for the conductance dependence, as $\alpha/e\vert\Sigma\vert \propto G^{-\gamma}$ with $\gamma = 1.2$. For  pH 10, the parameter $\alpha$ is found to be independent of $G$. {The origin of this change in exponent for higher pH is not understood at this level of analysis.}

{ {\it  Discussion --}
In conclusion we have presented a combined study of the ionic conduction and its fluctuations inside individual carbon nanotubes.
We have shown that in the case of CNT the conductance exhibits a power law behavior at low salinity, with an exponent close to 1/3 versus the salt concentration. This is in contrast to BN nanotubes which exhibits a constant surface conductance. We have been able to rationalize this behaviour in terms of a model accounting for hydroxide adsorption at the carbon surface. 
Similar process are likely to be at the origin of a similar observation in smaller single wall carbon nanotubes \cite{Strano,Lindsay2011,Forniasero}.

This analysis of the conductance is particularly useful in order to get insights into the ionic noise through CNTs.
The measurements of fluctuations in ionic conductance show an intimate link with the properties of the fluid-solid interface, as highlighted by a collapse of the low frequency noise amplitude once rescaled by the surface charge. 
This demonstrates that surface effects play a key role in the low frequency behavior of ionic transport. While the role of the surface charge on current fluctuation had been pointed for the high frequency regime \cite{Golovchenko}, such a link is shown here for the first time for the dominant low frequency component. The physical origin of such $1/f$ component to the noise remains largely mysterious. The scaling behavior measured in CNTs suggest new leads for the understanding of this slow dynamics in terms of the dynamics of adsorption processes at the confining surface. Work along these lines is in progress.\\
%
}

\acknowledgements{ LB aknowledges funding from the European Union's FP7 Framework Programme / ERC Advanced Grant Micromegas,  AS aknowledges funding from the European Union's H2020 Framework Programme / ERC Starting Grant agreement number 637748 - NanoSOFT. 
Authors thank D. Stein, B. Rotenberg, R. Netz and A. Michaelides for many fruitful discussions. 
}


\end{document}